\documentclass[aps,preprint]{revtex4}%
\usepackage{amsfonts}
\usepackage{amsmath}
\usepackage{amssymb}
\usepackage{subfigure}
\usepackage{graphicx}%
\usepackage{color}

\begin{document}
\title{Thermodynamics with pressure and volume of black holes based on two assumptions under scalar field scattering}
\author{Benrong Mu$^{a,b}$}
\email{benrongmu@cdutcm.edu.cn}

\author{Jing Liang$^{a,b}$}
\email{jingliang@stu.scu.edu.cn}

\author{Xiaobo Guo$^{c}$}
\email{guoxiaobo@czu.edu.cn}

\affiliation{$^{a}$ Physics Teaching and Research section, College of Medical Technology,
Chengdu University of Traditional Chinese Medicine, Chengdu, 611137,
PR China}
\affiliation{$^{b}$Center for Theoretical Physics, College of Physics, Sichuan University, Chengdu, 610064, PR China}
\affiliation{$^{c}$Mechanical and Electrical Engineering School, Chizhou University, Chizhou, 247000, PR China}

\begin{abstract}
Recently, a new assumption was proposed in [Phys. Rev. D 100, no.10, 104022 (2019)]. This assumption considers that the energy of the particle changes the enthalpy of the black hole after throwing the particle into the black hole. Using the energy-momentum relation, the results show that the second law of thermodynamics of the black hole is valid in extended phase space. In this paper, we discuss the validity of the laws of thermodynamics and the stability of the horizon of the charged AdS black hole by scalar field scattering under two assumptions, i.e., the energy flux of the scalar field $dE$ changes the internal energy of the black hole $dU$ and the energy flux of the scalar field $dE$ changes the enthalpy of the black hole $dM$.
\end{abstract}

\maketitle
\tableofcontents{}

\bigskip{}

\section{Introduction}
Since Bardeen et al. discovered that the black hole parameters satisfy equations similar to the laws of
thermodynamics \cite{intro-Bardeen:1973gs}, black hole mechanics was gradually replaced
by black hole thermodynamics. Black hole thermodynamics has gradually attracted attention.

For a RN-AdS black hole, the first law of black hole usually is
\begin{equation}
dM=TdS+\Phi dQ,
\label{eqn:1st1}
\end{equation}
where $M$ is the mass, $T$ is the Hawking temperature of the black hole, $S$ is the entropy, $\varPhi$ is the electric potential
and $Q$ is the electric charge. It is worth noting that there is no $PdV$ term in Eq. $\left(\ref{eqn:1st1}\right)$. The $PdV$ term in the
context of black hole space-time was eventually introduced in Ref. \cite{intro-Dolan:2012jh, intro-Kastor:2009wy}. The pressure $P$ was considered to be related to the $\varLambda$ \cite{intro-Dolan:2011xt,intro-Kubiznak:2012wp,intro-Cvetic:2010jb,intro-Caceres:2015vsa,intro-Hendi:2012um,intro-Pedraza:2018eey}.
Interpreting $M$ as the enthalpy, the first law of black hole is modified to
\begin{equation}
dM=TdS+VdP+\varPhi dQ.
\label{eqn:1st 2}
\end{equation}
In the above equation, $V$ is the volume of the black hole and is defined as the thermodynamic
conjugate to the pressure $P$. The idea that $\varLambda$ is considered as a thermodynamic variable has been considered by many authors \cite{intro-Henneaux:1984ji, intro-Henneaux:1989zc, intro-Teitelboim:1985dp, intro-Wang:2006eb, intro-Sekiwa:2006qj, intro-LarranagaRubio:2007ut, intro-Wang:2006bn, intro-Azreg-Ainou:2014lua, intro-Lemos:2018cfd}. In general, $M$ is related to the terms of the internal energy $U$ and $PV$ of the black hole
\begin{equation}
M=U+PV.
\end{equation}

The laws of thermodynamics is defined at the horizon of a black hole. The event horizon conceals the curvature singularity from
the outside observers \cite{intro-Penrose:1964wq}. When the horizon is destroyed, the singularity becomes a naked singularity. At the naked singularity, all the
laws of physics break down. Therefore, the horizon is assumed to be stable according to the weak cosmic censorship
conjecture (WCCC). Although it is generally accepted that this conjecture applies to black holes, its validity should be tested, since there is no general procedure to prove it. The validity of the WCCC can be tested by adding a particle with
energy and charge into the black hole \cite{intro-Sorce:1974dst}. If the horizon is destroyed after the particle absorption, the singularity becomes the naked
singularity and the WCCC is invalid. If the horizon exists, the singularity is surrounded by the horizon and the WCCC is satisfied. By this way, the stability of the horizon has been investigated in five-dimensional charged topological black holes \cite{intro-Chen:2020zps}, charged anti-de Sitter black holes \cite{intro-Chen:2019pdj}, generic charged NLED black hole \cite{intro-Wang:2019dzl}, Lanczos-Lovelock gravity \cite{intro-Jiang:2020alh} and Einstein-Born-Infeld black holes \cite{intro-He:2019mqy}. The effects of quantum gravity on black hole thermodynamics and the WCCC in the framework of GUP were studied in Ref. \cite{intro-Mu:2019bim}. The effect of quantum gravity
on the second law of thermodynamics and the WCCC of the RN rainbow black hole were discussed in \cite{intro-Gim:2019dvs}. The overcharging problem of a charged Taub-NUT black hole was discussed in \cite{intro-Feng:2020tyc}. Other studies on the stability of the event
horizon via particle absorption are referred to \cite{intro-Liang:2020uul, intro-He:2019fti, intro-Ying:2020bch, intro-Hu:2020lkg, intro-He:2019kws, intro-Wang:2019jzz, intro-Zeng:2019jta, intro-Gwak:2017kkt, intro-Isoyama:2011ea}.

Another important way to investigate the stability of the horizon is replacing a test
particle with a test field \cite{intro-Sorce:2017dst}. This way was first proposed by Semiz \cite{intro-Semiz:2005gs}. The validity of WCCC is determined by whether the event horizon exists after scalar field scattering. Semiz found that the WCCC was not violated in the dyonic Kerr-Newman black holes when considering the interaction between a charged black hole and a complex scalar field. But in the Kerr and BTZ black holes and in the Dirac field, the result obtained is different \cite{intro-Duztas:2013wua, intro-Duztas:2016xfg, intro-Duztas:2014sga}. Based on Semiz's work, Gwak studied the validity of the WCCC in the Kerr (anti-)de Sitter black holes by the scattering
of a scalar field, and found the horizon was stable \cite{intro-Gwak:2018akg}. In his work, the energy
and angular momentum in a certain time interval were related to their fluxes, respectively. Gwak's work supports the WCCC and is consistent with the conclusions reached by Wald and Sorce \cite{intro-Wald:2018xxi, intro-Sorce:2017dst}. Other studies on the stability of the event horizon under scalar field scatting are referred to \cite{intro-Liang:2020hjz, intro-Mu:2020szg, intro-Yang:2020czk, intro-Jiang:2020btc, intro-Hong:2020zcf, intro-Yang:2020iat, intro-Bai:2020ieh, intro-Hollands:2019whz, intro-Gwak:2019rcz, intro-Hong:2019yiz, intro-Gwak:2019asi, intro-Chen:2019nsr, intro-Chen:2018yah, intro-Toth:2011ab}.

When studied the validity of the laws of thermodynamics and the stability of the horizon via particle absorption, the energy of the particle is assumed to change the
internal energy of the black hole if considered the $PV$ term in previous papers. As shown in Ref. \cite{intro-Gwak:2017kkt}, Gwak considered throwing a
charged particle with energy $E$ and charge $q$ into a black hole. Then he proved the standard argument $p^{r}=E-q\Phi$ by means of
the Hamilton-Jacobi equation, where $p^{r}$ is denotes the radial momentum of the particle. Moreover, he assumed that the energy of the particle changes the
internal energy of the black hole, i.e., $E=dU$. Under this assumption, the first law of thermodynamics is satisfied, however the second law of
thermodynamics is indefinite. Moreover, the extremal black hole stays extremal and the near-extremal black hole stays near-extremal. As shown in Ref. \cite{intro-Hu:2019lcy}, the other assumption is proposed, i.e., the infalling particle changes the enthalpy of the black hole by $E=dM$. Under this assumption, the first and second laws of thermodynamics are well satisfied. Besides, the horizon is stable for the extremal and near-extremal black holes.

In this paper, we analyse the validity of the laws of thermodynamics and the stability of the horizon for a RN-AdS
black hole by the scattering of the charged scalar field. After the scattering of scalar fields, the mass and charge of the black hole change. The variation of the charge of the black hole $dQ$ is related to the variation of the electric charge flux of the scalar field $de$. In the previous papers, the variation of the internal energy of a black hole $dU$ is related to the variation of the energy flux of the scalar field $dE$. In this paper, we not only discuss the variation of the black hole under this assumption, but also study the validity of laws of thermodynamics and stability of the horizon when the variation of the enthalpy of the black hole $dM$ is related to the variation of the energy flux of the scalar field $dE$. The remainder of our paper is organized as follows. In section \ref{sec:M}, the thermodynamic properties of RN-AdS black holes are briefly reviewed. In section \ref{sec:T}, the first and second laws of thermodynamics under two assumptions are studied. In section \ref{sec:WCCC}, the stability of the horizon is discussed under two assumptions. Section \ref{sec:con} is devoted to our discussion and conclusion.

\section{Thermodynamics with pressure and volume in AdS black holes}
\label{sec:M}
The Einstein-Maxwell action of the charged scalar field with the cosmological constant in the 4-dimensional space-time is \cite{intro-Gwak:2017kkt}
\begin{equation}
S=-\frac{1}{16\pi}\int d^{4}x\sqrt{-g}\left(R-F_{\mu\nu}F^{\mu\nu}-2\varLambda\right),
\end{equation}
where the Maxwell field strength $F_{\mu\nu}$ is
\begin{equation}
F_{\mu\nu}=\partial_{\mu}A_{\nu}-\partial_{\nu}A_{\mu}.
\end{equation}
The metric of the RN-AdS black hole in 4-dimensional space-time is written as
\begin{equation}
ds^{2}=-f(r)dt^{2}+\frac{1}{f(r)}dr^{2}+r^{2}\left(d\theta^{2}+sin^{2}\theta d\phi^{2}\right),
\end{equation}
with
\begin{equation}
f(r)=1-\frac{2M}{r}+\frac{Q^{2}}{r^{2}}+\frac{r^{2}}{l^{2}}.
\end{equation}
In the above equation, $M$ is the mass of the black hole, $Q$ is the charge of the black hole and $l$ is the radius of the AdS space-time. The AdS radius $l$ is proportional to the cosmological constant $\Lambda$ as $\Lambda=-\frac{3}{l^{2}}$. The thermodynamic properties is defined on the black hole's horizon $r_{+}$. The Hawking-temperature, Bekenstein-Hawking entropy and electric potential are written as
\begin{equation}
T=\frac{f^{\prime}(r_{+})}{4\pi}=\frac{1}{2\pi}(\frac{M}{r_{+}^{2}}-\frac{Q^{2}}{r_{+}^{3}}+\frac{r_{+}}{l^{2}}),
\label{eqn:T}
\end{equation}
\begin{equation}
S=\pi r_{+}^{2},
\label{eqn:S}
\end{equation}
\begin{equation}
\Phi=\frac{Q}{r_{+}}.
\label{eqn:Phi}
\end{equation}
In extended phase space, the cosmological constant is regarded as a thermodynamic variable, and exhibits a fairly consistent behavior with other thermodynamic variables. The cosmological constant $\Lambda$ plays the role of pressure, and the thermodynamic volume $V$ is defined as the conjugate variable of the pressure. The pressure and volume can be expressed as
\begin{equation}
P=-\frac{\Lambda}{8\pi}=\frac{3}{8\text{\ensuremath{\pi}}l^{2}},V=\left(\frac{\partial M}{\partial P}\right)_{S,Q}=\frac{4\text{\ensuremath{\pi}}r_{+}^{3}}{3}.
\label{eqn:PV}
\end{equation}
When the pressure term is considered in the laws of thermodynamics, the key difference is
that the mass now plays the roles of enthalpy in the first law of thermodynamics \cite{intro-Kastor:2009wy, intro-Cvetic:2010jb}. Therefore, the first law of thermodynamics in extended phase space is \cite{intro-Dolan:2012jh, M-Dolan:2013ft}
\begin{equation}
dM=TdS+\Phi dQ+VdP.
\end{equation}
Enthalpy is related to the terms of the internal energy $U$ and $PV$ of the black hole
\begin{equation}
M=U+PV.
\end{equation}

\section{Thermodynamics under Charged Scalar Field}
\label{sec:T}
A black hole can obtain conserved quantities such as energy, momentum, and charge through the scattering of a scalar field. The conserved quantity that the black hole obtained by the scattering of a scalar field is given by the flux of the scattered scalar field. A black hole can change its state when it interacts with the external field.
From the flux, the variation of the black hole over an infinitesimal time interval can be estimated.
As shown in Ref. \cite{intro-Gwak:2019asi}, fluxes of energy and electric charge of the scalar field are
\begin{equation}
\begin{aligned}
&\frac{dE}{dt}=\omega(\omega-q\Phi)r_{+}^{2},\\
&\frac{de}{dt}=q\left(\omega-q\Phi\right)r_{+}^{2}.\\
\end{aligned}
\end{equation}
In Ref. \cite{intro-Gwak:2019asi}, Gwak related the electric charge flux to the change in that of
the black hole. In following sections, laws of thermodynamics will be discussed separately under the assumption that the energy flux is related to internal energy and the assumption that the energy flux is related to enthalpy.

After the scalar field scattering, the state of a black hole changes. The outer horizon radius $r_{+}$ moves to $r_{+}+dr_{+}$ and function $f(r)$ moves from $f(M,Q,l,r_{+})$ to $f(M+dM,Q+dQ,l+dl,r_{+}+dr_{+})$. Assuming that the final state of the black hole is still a black hole. The initial state $f(M,Q,l,r_{+})$ and final state $f(M+dM,Q+dQ,l+dl,r_{+}+dr_{+})$ satisfy
\begin{equation}
f\left(M+dM,Q+dQ,l+dl,r_{+}+dr_{+}\right)=f\left(M,Q,l,r_{+}\right)+df=0.
\label{eqn:nor T f}
\end{equation}
The shift of $f(r)$ is
\begin{equation}
df=\frac{\partial f}{\partial M}|_{r=r_{+}}dM+\frac{\partial f}{\partial Q}|_{r=r_{+}}dQ+\frac{\partial f}{\partial l}|_{r=r_{+}}dl+\frac{\partial f}{\partial r}|_{r=r_{+}}dr_{+}=0,
\label{eqn:ext T df1}
\end{equation}
where
\begin{equation}
\begin{aligned}
& \frac{\partial f}{\partial M}|_{r=r_{+}}=-\frac{2}{r_{+}},\frac{\partial f}{\partial Q}|_{r=r_{+}}=\frac{2Q}{r_{+}^{2}},\\
& \frac{\partial f}{\partial l}|_{r=r_{+}}=-\frac{2r_{+}^{2}}{l^{3}},\frac{\partial f}{\partial r}|_{r=r_{+}}=4\text{\ensuremath{\pi}}T.\\
\end{aligned}
\label{eqn:ext T df}
\end{equation}
Then, we study the variation of a charged AdS black hole due to the inflow flux of the scattered scalar field base on two assumptions.
\subsection{Assumption A: the energy flux changes the internal energy of the black hole by $dE=dU$}

After scalar field scattering, the charge flux clearly corresponds to changing the charge of the black hole. In Ref. \cite{intro-Gwak:2019asi}, Gwak relates the energy flux to the internal energy. Hence
\begin{equation}
\begin{aligned}
&dU=dM-d(PV)=\omega(\omega-q\Phi)r_{+}^{2}dt,\\
&dQ=q(\omega-q\Phi)r_{+}^{2}dt,\\
\end{aligned}
\label{eqn:dUdQ}
\end{equation}
where $d(PV)=PdV-VdP$. Combing Eqs. $\left(\ref{eqn:ext T df1}\right)$, $\left(\ref{eqn:ext T df}\right)$ and  $\left(\ref{eqn:dUdQ}\right)$, we obtain
\begin{equation}
dr_{+}=\frac{2r_{+}l^{2}(\omega-q\Phi)^{2}}{4\text{\ensuremath{\pi}}Tl^{2}-3r_{+}}dt.
\label{eqn:T dr1}
\end{equation}
Therefore, the entropy variation is
\begin{equation}
dS=2\text{\ensuremath{\pi}}r_{+}dr_{+}=\frac{4\text{\ensuremath{\pi}}r_{+}^{2}l^{2}(\omega-q\Phi)^{2}}{4\text{\ensuremath{\pi}}l^{2}T-3r_{+}}dt.
\label{eqn:T dS1}
\end{equation}
Incorporating Eqs. $\left(\ref{eqn:T}\right)$, $\left(\ref{eqn:S}\right)$, $\left(\ref{eqn:Phi}\right)$ and $\left(\ref{eqn:T dS1}\right)$,
the energy relation in $\left(\ref{eqn:dUdQ}\right)$ becomes
\begin{equation}
dM=TdS+\Phi dQ+VdP,
\label{eqn:T dM}
\end{equation}
Thus, the first law of thermodynamics of the RN-AdS black hole is recovered by the scattering
of the scalar field when assuming that the energy flux changes the internal energy.

The second law of thermodynamics states that the increment of the entropy of a black hole in an irreversible process is always greater than zero. Since the scalar field scattering is an irreversible process, the entropy of the black hole is larger than the entropy before
the scalar field scattering. The validity of the second law of thermodynamics can be verified by the sign of $dS$ in Eq. $\left(\ref{eqn:T dS1}\right)$.

For the extremal black hole, $T=0$. Then, $dS$ has a negative value
\begin{equation}
dS=-\frac{4\text{\ensuremath{\pi}}r_{+}^{2}l^{2}(\omega-q\Phi)^{2}}{3r_{+}}dt<0.
\end{equation}
Thus, the entropy of the black hole decreases at least for the extremal case. Therefore, the second law of thermodynamics is violated when the energy flux is related to the internal energy under charge scalar field.

\begin{figure}[htb]
\centering
\includegraphics[scale=0.8]{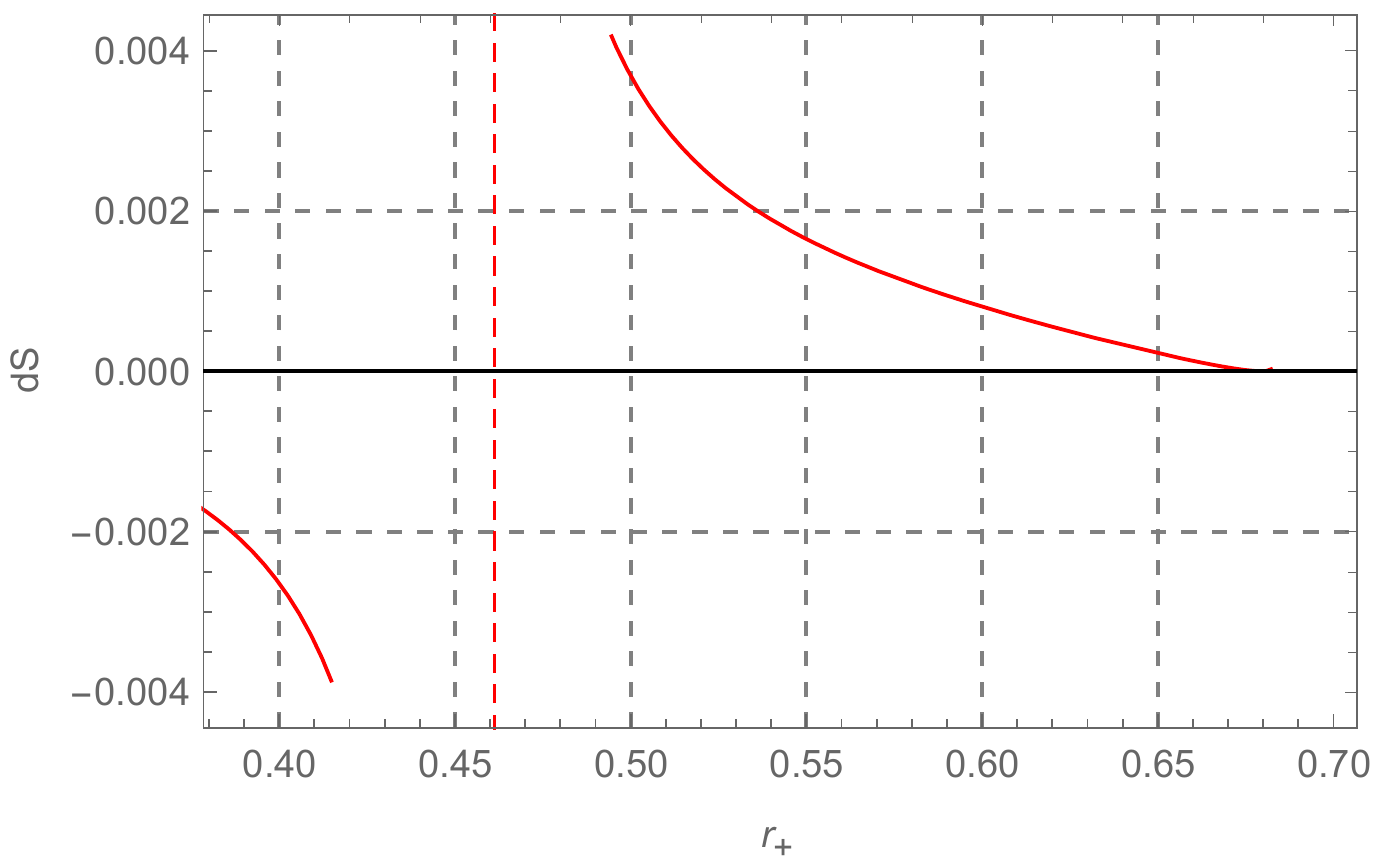}
\caption{The relation between $dS$ and $r_+$ which parameter values are $M=0.5$, $l=1$, $\omega=0.1$ and $dt=0.001$. When $r_+ > 0.385458$,
the value of temperature $T$ is greater than zero.}
\label{fig:dS1}
\end{figure}

We numerically analyze and plot Eq. $\left(\ref{eqn:T dS1}\right)$. as Fig. \ref{fig:dS1}. The imaginary line in Fig. \ref{fig:dS1} has a value of $r_+ = 0.9679$ is the extremal black hole horizon radius. As shown in Fig. \ref{fig:dS1}, the value of $dS$ is negative for $r_+$ less than 0.4612 and positive for $r_+$ greater than
0.4612. The range of values of $r_+$ in the Fig. \ref{fig:dS1} is [0.385458,0.70], when the value of $r_+$ is less than 0.385458, the value of $T$ is less than 0.

\subsection{Assumption B: the energy flux changes the enthalpy of the black hole by $dE=dM$}
In the previous section, the second law of thermodynamics is violated when assuming that the energy flux corresponds to the internal energy of the black hole. Then, we consider the validity of the laws of thermodynamics under another assumption. Based on Ref. \cite{intro-Hu:2019lcy}, the energy flux of the scalar field is assumed to change the black hole's enthalpy in this section.

When the energy flux is assumed to change the enthalpy of the black hole, the Eq. $\left(\ref{eqn:dUdQ}\right)$ becomes
\begin{equation}
\begin{aligned}
&dM=\omega(\omega-q\Phi)r_{+}^{2}dt,\\
&dQ=q(\omega-q\Phi)r_{+}^{2}dt.\\
\end{aligned}
\label{eqn:T dMQ}
\end{equation}
Repeating the same calculation in the previous subsection, we get the value of the $dr_+$ under this assumption, which is
\begin{equation}
dr_{+}=\frac{r_{+}(\omega-q\Phi)^{2}dt+r_{+}^{2}l^{-3}dl}{2\text{\ensuremath{\pi}}T}.
\label{eqn:T dr1}
\end{equation}
Besides, considering Eq. $\left(\ref{eqn:S}\right)$, the change of the black hole entropy is written as
\begin{equation}
dS=2\text{\ensuremath{\pi}}r_{+}dr_{+}=\frac{r_{+}^{2}(\omega-q\Phi)^{2}dt+r_{+}^{3}l^{-3}dl}{T}.
\label{eqn:T dS2}
\end{equation}
Substituting Eqs. $\left(\ref{eqn:T}\right)$, $\left(\ref{eqn:S}\right)$, $\left(\ref{eqn:Phi}\right)$ and $\left(\ref{eqn:T dS2}\right)$
into Eq. $\left(\ref{eqn:T dMQ}\right)$, we have
\begin{equation}
dM=TdS+\Phi dQ+VdP,
\end{equation}
which is the first law of thermodynamics.

As before, we verify the validity of the second law of thermodynamics by discuss the sign of $dS$ in Eq. $\left(\ref{eqn:T dS1}\right)$. It is obviously that the value of $dS$ is always positive, though approaching zero
in the extremal limit. Thus, the second law of thermodynamics is always satisfied when we assume that the energy flux changes the enthalpy of the black hole.

\begin{figure}
\centering
\includegraphics[scale=0.8]{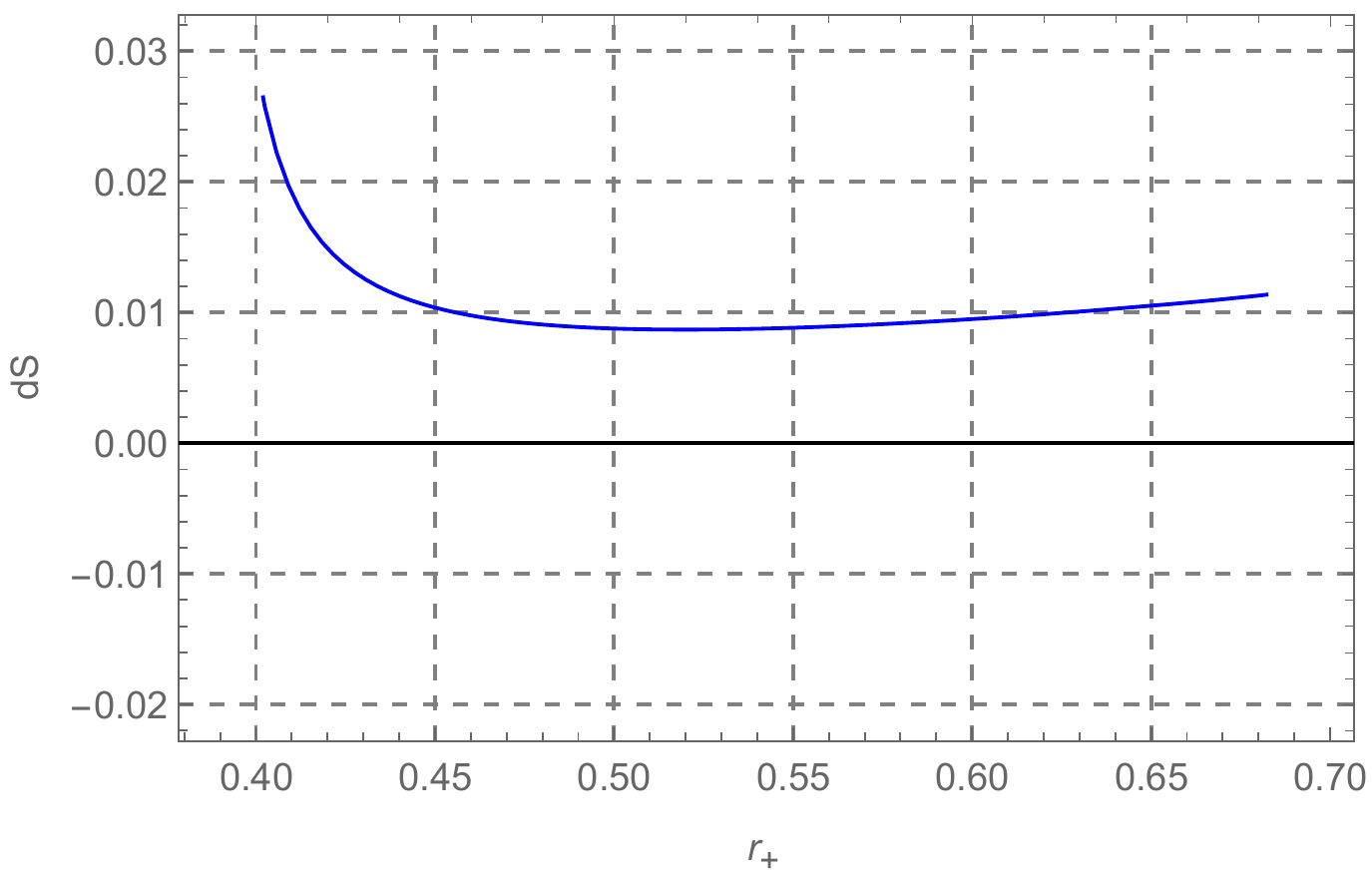}
\caption{The relation between $dS$ and $r_+$ which parameter values are $M=0.5$, $l=1$, $\omega=0.1$ and $dt=0.001$. When $r_+ > 0.385458$, the value of temperature $T$ is greater than zero.}
\label{fig:dS2}
\end{figure}

As shown in Fig. \ref{fig:dS2}, the value of $dS$ is always positive for $r_+$ greater than 0.385458. The range of values of $r_+$ in the Fig. \ref{fig:dS2} is [0.385458,0.70], when the value of $r_+$ is less than 0.385458, the value of $T$ is negative.
\section{Stability of Horizon}
\label{sec:WCCC}
An important feature of a black hole is its event horizon. The horizon divides the inside and outside of the black hole.
In addition, thermodynamic variables are defined at the event horizon. After the scalar field scattering, the mass and charge of
the black hole change, and the outer and inner horizons also change. In this section, we consider the stability of the event horizon
after scalar field scattering. The stability of the event horizon is determined by the sign of the minimum value of $f(r)$. If the minimum value of $f(r)$ is not greater than
zero, there exists solutions corresponding to horizons. If the minimum value of $f(r)$ is positive, there is no solution represents
the horizon. Thus, the stability of horizon can be verified by the sign of $f(r)$ after the scalar field scatting.

We assume that the initial state is a near-extremal black hole. When the fluxs of the scalar field enters the black hole, the mass,
charge and the AdS radius of the black hole change from $(M,Q,l)$ to $(M+dM,Q+dQ,l+dl)$. Meanwhile, the location of the event
horizon and the minimum value change to $(r_{min}+dr_{min}, r_++dr_+)$. Corresponding, the minimum value of $f(r)$ at
$(r_{min}+dr_{min})$ is
\begin{equation}
\begin{aligned}
& f(M+dM,Q+dQ,l+dl,r+dr)|_{r=r_{min}}\\
& =\delta+\frac{\partial f}{\partial M}|_{r=r_{min}}dM+\frac{\partial f}{\partial Q}|_{r=r_{min}}dQ+\frac{\partial f}{\partial l}|_{r=r_{min}}dl+\frac{\partial f}{\partial r}|_{r=r_{min}}dr_{min},\\
\end{aligned}
\label{eqn:W f1 eqn}
\end{equation}
where
\begin{equation}
\begin{aligned}
& \frac{\partial f}{\partial M}|_{r=r_{min}}=-\frac{2}{r_{min}},\frac{\partial f}{\partial Q}|_{r=r_{min}}=\frac{2Q}{r_{min}^{2}},\\
& \frac{\partial f}{\partial l}|_{r=r_{min}}=-\frac{2r_{min}^{2}}{l^{3}},\frac{\partial f}{\partial r}|_{r=r_{min}}=0.
\end{aligned}
\label{eqn:W df}
\end{equation}
Inserting Eq. $\left(\ref{eqn:W df}\right)$ into $\left(\ref{eqn:W f1 eqn}\right)$, we obtain
\begin{equation}
\begin{aligned}
& f(M+dM,Q+dQ,l+dl,r+dr)|_{r=r_{min}}=\delta-\frac{2}{r_{min}}dM+\frac{2Q}{r_{min}^{2}}dQ-\frac{2r_{min}^{2}}{l^{3}}dl.\\
\end{aligned}
\label{eqn:W f min}
\end{equation}
Then, we investigate the value of $f(M+dM,Q+dQ,l+dl,r+dr)|_{r=r_{min}}$ under two assumptions.
\subsection{Assumption A: the energy flux changes the internal energy of the black hole by $dE=dU$}
When the energy and charge fluxes are related to the internal energy and electric charge, the changes in internal energy and electric charge are given as
$dU=dM-d(PV)=\omega(\omega-q\Phi)r_{+}^{2}dt$ and $dQ=q(\omega-q\Phi)r_{+}^{2}dt$. Then, Eq. $\left(\ref{eqn:W f min}\right)$
changes into
\begin{equation}
\begin{aligned}
& f(M+dM,Q+dQ,l+dl,r+dr)|_{r=r_{min}}\\
&=\delta-\frac{2r_{+}^{2}(\omega-q\Phi)^{2}}{r_{min}^{2}}\left(r_{min}-r_{+}\right)dt-\frac{3r_{+}}{2\pi r_{min}l^{2}}dS-\frac{2\left(r_{min}^{3}-r_{+}^{3}\right)}{r_{min}l^{3}}dl.\\
\end{aligned}
\label{eqn:W f11}
\end{equation}
When it is an extreme black hole at the initial state, $r_{min}=r_{+}$, $T=0$ and $\delta=0$. Then, we obtain
\begin{equation}
f(M+dM,Q+dQ,l+dl,r+dr)|_{r=r_{min}}=\frac{2r_{+}^{2}(\omega-q\Phi)^{2}}{r_{min}}dt,
\end{equation}
where $dt$ is an infinitesimal scale and is set as $dt\sim\epsilon$, where $0<\epsilon\ll1$. Then, we skip to write the first order of $\epsilon$ and the final state of the black hole has
\begin{equation}
f(M+dM,Q+dQ,l+dl,r+dr)|_{r=r_{min}}=0.
\end{equation}
This implies that the horizon is still exists at the final state of the extremal black hole.

When it is the near-extremal black hole at the initial state, $r_{min}$ and $r_+$ do not coincide. We suppose that $r_{+}=r_{min}+\epsilon$, where $0<\epsilon\ll1$.
Then, the minimum value of $f(r)$ at the final state is given by
\begin{equation}
\begin{aligned}
& f(M+dM,Q+dQ,l+dl,r+dr)|_{r=r_{min}}\\
& =\delta+\frac{2r_{+}^{2}(\omega-q\Phi)^{2}}{r_{min}^{2}}\epsilon dt-\frac{3\left(r_{min}+\epsilon\right)}{2\pi r_{min}l^{2}}dS+\frac{6r_{min}}{l^{3}}\epsilon dl,
\end{aligned}
\end{equation}
where $dt$ is an infinitesimal scale and is set as $dt\sim\epsilon$. If the initial black hole is near extremal, we have $dS\sim\epsilon$, $dt\sim\epsilon$ and $dl\sim\epsilon$.
Thus, the above equation becomes
\begin{equation}
f(M+dM,Q+dQ,l+dl,r+dr)|_{r=r_{min}} \thickapprox \delta \leq 0.
\end{equation}
Therefore, the horizon stably exists at the final state of the near-extremal black hole.
\subsection{Assumption B: the energy flux changes the enthalpy of the black hole by $dE=dM$}
When the energy and charge fluxes are related to the enthalpy and electric charge, the changes in enthalpy and electric charge are given as
$dM=\omega(\omega-q\Phi)r_{+}^{2}dt,dQ=q(\omega-q\Phi)r_{+}^{2}dt.$ Then, Eq. $\left(\ref{eqn:W f min}\right)$ is written as
\begin{equation}
\begin{aligned}
& f(M+dM,Q+dQ,l+dl,r+dr)|_{r=r_{min}}\\
&=\delta-\frac{2r_{+}^{2}(\omega-q\Phi)}{r_{min}^{2}}\left(r_{min}\omega-r_{+}q\Phi\right)dt-\frac{2r_{min}^{2}}{l^{3}}dl.\\
\end{aligned}
\label{eqn:W f12}
\end{equation}
For the extremal black hole, $r_{min}=r_{+}$, $T=0$ and $\delta=0$. Then the above equation becomes
\begin{equation}
f(M+dM,Q+dQ,l+dl,r+dr)|_{r=r_{min}}=-2r_{+}(\omega-q\Phi)^{2}dt-\frac{2r_{min}^{2}}{l^{3}}dl<0.
\end{equation}
For the near-extremal black hole, $r_{min}$ and $r_+$ do not coincide and $\delta<0$. We can suppose that $r_{+}=r_{min}+\epsilon$, where $0<\epsilon\ll1$. Then we obtain
\begin{equation}
\begin{aligned}
&f(M+dM,Q+dQ,l+dl,r+dr)|_{r=r_{min}}\\
&=\delta-\frac{2\left(r_{min}+\epsilon\right)^{2}(\omega-q\Phi)^{2}}{r_{min}}dt+\frac{2\left(r_{min}+\epsilon\right)^{2}(\omega-q\Phi)q\Phi}{r_{min}^{2}}\epsilon dt-\frac{2r_{min}^{2}}{l^{3}}dl,\\
\end{aligned}
\end{equation}
where $dt$ is an infinitesimal scale and is set $dt\sim\epsilon$. Therefore, it is obviously that the value of $f(M+dM,Q+dQ,l+dl,r+dr)|_{r=r_{min}}$ is less than zero. Hence, the event horizon remains stable after scattering of the scalar field.

\section{Discussion and Conclusion}
\label{sec:con}
In this paper, the laws of thermodynamics and the stability
of horizon of a RN-AdS black hole in 4-dimensional space-time were calculated and discussed via
scalar field scattering in extended phase space. Two assumptions were considered, namely the energy flux of the scalar field changes the internal energy of the black hole and the energy flux of the scalar field changes
the enthalpy of the black hole. The first law of thermodynamics is recovered under both assumptions. The second law of thermodynamics is violated when the energy flux is assumed to correspond to the internal energy of the black hole, and is valid when the energy flux is assumed to correspond to the enthalpy of the black hole. Besides, the horizon is stable under both assumptions.

\begin{table}[htb]
\begin{centering}
\begin{tabular}{|p{1.0in}|p{1.2in}|p{1.2in}|p{1.2in}|p{1.2in}|}
  \hline  &\multicolumn{2}{|c|}{Normal phase space}  &\multicolumn{2}{|c|}{Extended phase space}\\
  \hline &E=dU &E=dM &E=dU &E=dM\\
  \hline 1st law  &$dM=TdS+\Phi dQ.$ &$dM=TdS+\Phi dQ.$ &$dM=TdS+\Phi dQ+VdP.$ &$dM=TdS+\Phi dQ+VdP.$\\
  \hline 2nd law  &Satisfied.

  $dS=\frac{p^{r}}{T}$. &Satisfied.

  $dS=\frac{p^{r}}{T}$. &Indefinite.

  $dS=\frac{4\pi l^{2}p^{r}}{4\pi l^{2}T-3r_{+}}.$ &Satisfied.

  $dS=\frac{p^{r}+r_{+}^{3}l^{-3}dl}{T}.$\\
  \hline The stability of horizon
  &The horizon still exists for the extremal and near-extremal black holes.
  &The horizon still exists for the extremal and near-extremal black holes.
  &The horizon still exists for the extremal and near-extremal black holes.
  &The horizon still exists for the extremal and near-extremal black holes.\\
  \hline
\end{tabular}
\par\end{centering}
\caption{{\footnotesize{}{}{}{}Results for the first and second laws of
thermodynamics and the stability of horizon for RN-AdS black hole via particle absorption.}}
\label{tab:result2}
\end{table}

The validity of the laws of thermodynamics and the stability of the horizon in extended phase space can be discussed in two ways, namely particle absorption and scalar field scattering. As shown in Ref. \cite{intro-Gwak:2017kkt}, after throwing
a particle with energy $E$ and charge $q$ into the black hole, the conserved quantities of the particle is absorbed into the conserved quantities of the black hole at the outer horizon $r_+$. The energy relationship between the conserved quantities and the momenta at the outer horizon is $E=q\Phi+p^{r}$, where $p^{r}$ is the radial momentum of the particle. In Ref. \cite{intro-Gwak:2017kkt}, the energy of the particle is assumed to change the internal energy of the black hole. Then the relation $d(M-PV)=\Phi dQ+p^{r}$ obtained. Under this assumption, for a RN-AdS black hole, the first law of thermodynamics is satisfied but the second law of thermodynamics is violated. Besides, the horizon is stable and the WCCC is valid. In Ref. \cite{intro-Hu:2019lcy}, another assumption was proposed. The first and second laws of thermodynamics of a RN-AdS black hole were discussed via particle absorption under two assumptions, i.e., the energy of the particle is assumed to correspond to the internal energy and is assumed to correspond to the enthalpy. When the energy of the particle is assumed to change the internal energy of the black hole, i.e., $dU=\Phi dQ+p^{r}$, the first law of thermodynamics is satisfied but the second law of thermodynamics is violated. When the energy of the particle is assumed to change the enthalpy of the black hole, i.e., $dM=\Phi dQ+p^{r}$, the first and second laws of thermodynamics are both valid. The results of the black hole under two assumptions in normal phase space are same, since the mass can be regarded as the internal energy in normal phase space. Based on Ref. \cite{intro-Hu:2019lcy}, we further calculated and discussed the stability of the horizon of the RN-AdS black hole under two assumptions via particle absorption. The conclusions are summarized in Table \ref{tab:result2}. Besides, the laws of thermodynamics and weak cosmic censorship conjecture of the conformal anomaly corrected AdS black hole were discussed under assumption $E=dM$ both in normal and extended phase spaces via particle absorption \cite{con-Li:2020dnc}.

\begin{table}[htb]
\begin{centering}
\begin{tabular}{|p{0.7in}|p{1.2in}|p{1.2in}|p{1.5in}|p{1.7in}|}
  \hline  &\multicolumn{2}{|c|}{Normal phase space}  &\multicolumn{2}{|c|}{Extended phase space}\\
  \hline &dE=dU &dE=dM &dE=dU &dE=dM\\
  \hline 1st law  &$dM=TdS+\Phi dQ.$ &$dM=TdS+\Phi dQ.$ &$dM=TdS+\Phi dQ+VdP.$ &$dM=TdS+\Phi dQ+VdP.$\\
  \hline 2nd law  &Satisfied.

  $dS=\frac{r_{+}^{2}(\omega-q\omega)^{2}}{T}dt$. &Satisfied.

  $dS=\frac{r_{+}^{2}(\omega-q\omega)^{2}}{T}dt$. &Indefinite.

  $dS=\frac{4\text{\ensuremath{\pi}}r_{+}^{2}l^{2}(\omega-q\Phi)^{2}}{4\text{\ensuremath{\pi}}l^{2}T-3r_{+}}dt.$ &Satisfied.

  $dS=\frac{r_{+}^{2}(\omega-q\Phi)^{2}dt+r_{+}^{3}l^{-3}dl}{T}.$\\
  \hline The stability of horizon
  &The horizon still exists for the extremal and near-extremal black holes.
  &The horizon still exists for the extremal and near-extremal black holes.
  &The horizon still exists for the extremal and near-extremal black holes.
  &The horizon still exists for the extremal and near-extremal black holes.\\
  \hline
\end{tabular}
\par\end{centering}
\caption{{\footnotesize{}{}{}{}Results for the first and second laws of
thermodynamics and the stability of horizon for RN-AdS black hole via scalar field scattering.}}
\label{tab:result}
\end{table}

Another important way to investigate the validity of the laws of thermodynamics and the stability of the horizon is scalar field scattering. According to Ref. \cite{intro-Hu:2019lcy}, the energy of the particle is assumed to change the internal energy or the enthalpy of the black hole. Therefore, we supposed the energy of the scalar field can also be assumed to change internal energy or enthalpy of the black hole. As shown in Ref. \cite{intro-Gwak:2017kkt}, the carried energy $E$ and electric charge $Q$ of the scalar field are given as their fluxes at the outer horizon. Fluxes of the scalar field will infinitesimally change the corresponding properties of the black hole during the scalar field scattering. The electric charge flux corresponds to the change of the black hole. In most of the previous papers \cite{intro-Liang:2020hjz, intro-Mu:2020szg, intro-Yang:2020czk, intro-Jiang:2020btc, intro-Hong:2020zcf}, the energy flux is assumed to correspond to internal energy of the black hole, i.e., $dE=dU$. Under this assumption, the second law of thermodynamics for black holes is violated in extended phase space.
Based on Ref. \cite{intro-Hu:2019lcy}, we considered the other assumption. Under this assumption, the energy flux is assumed to change the enthalpy of the black hole instead of the internal energy of the black hole, i.e., $dE=dM$. Then, the second law of thermodynamics of the black hole was found to be valid. In addition, the first law of thermodynamics and the stability of the horizon under this assumption have the same results as the previous one. In normal phase space, the mass can be regarded as the internal energy. Therefore, the results of the black hole under two assumptions are same in normal phase space. The conclusions are summarized in Table \ref{tab:result}.

In Ref. \cite{intro-Gwak:2019asi}, laws of thermodynamics and the WCCC in charged AdS black holes with thermodynamic pressure and volume by the scattering of the charged scalar field in four and higher dimensions are investigated. During the discussion, the variation of internal energy and charge of the black hole are related to the energy and charge flux of the scalar field. Then, the first law of thermodynamics is recovered, although the second law of thermodynamics is violated in extremal and near-extremal black holes. Furthermore, the WCCC remains valid, since the extremality of a black hole is invariant despite changes in internal energy and electric charge. Although in this paper, we have also studied the laws of thermodynamics and the stability of the horizon of RN-AdS black holes in four dimensions by scalar field scattering, we consider another assumption based on Ref. \cite{intro-Hu:2019lcy}. Under this assumption, the charge of the black hole still corresponds to the charge flux of the scalar field, while the change of the enthalpy of the black hole corresponds to the energy flux of the scalar field. Then, we found that both the first and second laws of thermodynamics are valid, as with the results obtained by particle absorption. The laws of thermodynamics in charged AdS black holes are studied under two assumptions via particle absorption in Ref. \cite{intro-Hu:2019lcy}. Furthermore, the stability of horizon was also discussed under two assumptions by scalar field scattering in this paper. And we found the WCCC is valid and the horizon is stable for extremal and near-extremal black holes.

Therefore, it can be seen that the results give us a new way for subsequent studies, where the energy flux of the scalar field changes the enthalpy of the black hole. We hope to see further studies of black hole thermodynamics under these two assumptions in the future.

\begin{acknowledgments}
We are grateful to Deyou Chen, Peng Wang, Jun Tao and Haitang Yang for useful discussions. This work is supported in part by NSFC (Grant No. 11747171), Natural Science Foundation of Chengdu University of TCM (Grants nos. ZRYY1729 and ZRYY1921), Discipline Talent Promotion Program of /Xinglin Scholars(Grant no.
QNXZ2018050), the key fund project for Education Department of Sichuan (Grant
no. 18ZA0173), the open fund of State Key Laboratory of Environment-friendly
Energy Materials of Southwest University of Science and Technology(Grant no.
17kffk08) and Special Talent Projects of Chizhou University (Grant no. 2019YJRC001).
\end{acknowledgments}

\end{document}